# Pattern Separation in a Spiking Neural Network of Hippocampus Robust to Imbalanced Excitation/Inhibition

**Faramarz Faghihi[1], Homa Samani[2], Ahmed A.Moustafa[3],***

1. Department of Computational Neuroscience, Bernstein Center for Computational Neuroscience, III. Institute of Physics - Biophysics, Georg-August-Universität, Göttingen, Germany
2. Department of Biology, Isfahan University, Isfahan, Iran
3. *School of Social Sciences and Psychology and Marcs Institute for Brain and Behavior, Western Sydney University, Sydney, NSW, Australia (*corresponding author*). Email: a.moustafa@westernsydney.edu.au

**Abstract**

Efficient pattern separation in dentate gyrus plays an important role in storing information in the hippocampus. Current knowledge of the structure and function of the hippocampus, entorhinal cortex and dentate gyrus, in pattern separation are incorporated in our model. A three-layer feedforward spiking neural network inspired by the rodent hippocampus an equipped with simplified synaptic and molecular mechanisms is developed. The aim of the study is to make an spiking neural network capable of pattern separation in imbalanced excitation/inhibition ratios caused by different levels of stimulations or network damage. This work present a novel theory on the cellular mechanisms of robustness to damges to synapses and connectivity of neurons in dentate gyrus that results in imbalanced excitation-inhibition activity of neurons. This spiking neural network uses simplified molecular and cellular hypothetical mechanisms and demonstrates efficient storing of information in different levels of stimulation and can be implemented in cognitive robotics.

**Keywords:** spiking neural networks, memory, dentate gyrus, pattern separation, imbalanced network, back-propagation, hippocampus

## Introduction

Computational scientists use neuroscience information to develop novel artificial systems. The main challenge of these research is the limited understanding of the neurobiology of cognitive functions and the parameters involved in neural systems. Neural systems' architecture and mesoscopic information of neurons' structure and their electrophysiological features play critical roles in the cognitive functions including different memory capabilities of rodents.

Progress in our understanding of the molecular, cellular and network architecture features of animals' brain and their cognitive capabilities have inspired engineers to develop intelligent machines and cognitive architectures [1, 2]. Currently, artificial intelligence has aimed to develop novel inspired artificial memories and learning systems [3, 4].

The episodic memory system in rodents [5, 6] is associated with the medial temporal lobe including the hippocampus [7]. Moreover, different kinds of memory systems are supported by the interaction of the hippocampus and neocortex [8]. The hippocampus encodes novel information as one-shot learning that is gradually consolidated to the neocortex during sleep [9, 10].

The role of the hippocampus in the acquisition and retention of episodic memory and spatial memory has been extensively studied [11, 12]. The hippocampus mediates many cognitive functions, such as spatial encoding [13]. But the memory trace consists of episodic-like events are encoded in parallel by the hippocampus and neocortex. Hippocampal-neocortical interaction theory attempts to find neural circuits and cellular mechanisms in hippocampus and neocortex that have been identified fundamental to memory formation and retrieval [14].

Spiking neural networks use biologically-realistic models of neurons to perform neural computations. SNNs have many applications in pattern recognition methods [15], image classification [16], modeling of EEG data [17], robotic navigation [18] and computer vision [19, 20]. Moreover, they have been used in modeling biological neural systems to predict systems' behavior. For examples, in simulating of organization of model of the basal ganglia circuitry [21], in modeling of hippocampus circuits [22], in modeling of olfactory system [23], in modeling of dopaminergic systems [24] and in modeling of Storage and retrieval of different memories in hippocampus [25, 26].

Cortical neurons receive thousands of excitatory and inhibitory synaptic inputs. Excitation-inhibition (E/I) dynamics in neural circuits must be carefully balanced to achieve temporally precise spiking [27]. Balancing the excitatory and inhibitory currents is crucial to keep the neurons at a functional dynamic range allowing them to elicit action potentials in response to their inputs [28]. In addition, the excitatory and inhibitory balance may be a fundamental mechanism that determines efficient coding in the auditory cortex [29]. In addition, imbalances in excitatory and inhibitory brain circuitry may be a main cause of neurodevelopmental cognitive disorders [30]. The importance of balanced excitation-inhibition in SNNs with applications in developing efficient brain-like neuromorphic architectures has been studied [31, 32]. Recent modeling studies have shown the stabilizing effect of inhibitory plasticity on network dynamics by a dynamic balance of network currents between excitation and inhibition [33, 34]. In [34], a spiking neural model is trained to discriminate patterns as an unsupervised learning method. The pattern discrimination by this network is impaired in case of imbalanced excitation-inhibition activity of neurons. Imbalanced excitation-inhibition ratio may be emerged as the consequence of change in connectivity of neurons [44] or impairments in neurotransmitter release in synapses [35]. The neural systems are expected to be equipped with cellular mechanisms to rebalance excitation-inhibition activities of neural populations in response to sensory inputs.

The role of connectivity of inhibitory neurons in balancing excitation-inhibition cortical networks has been theoretically studied [36]. Another parameter that may influence maintenance of the excitation-inhibition balance is inhibition plasticity [37, 38] and may be established in an experience-dependent manner by synaptic plasticity of inhibitory interneurons. This mechanism may provide an explanation for sparse firing pattern observed in neocortex in response to stimulation [39]. Molecular mechanisms involved in synaptogenesis (formation of new synapses between neurons) may influence the excitation-inhibition ratios in CNS (Central Nervous System) [40].

The hippocampus of rodents is composed as Dentate Gyrus (DG), CA3, CA2, CA1, Subuculum (sub) and Entorhinal Cortex (EC). The DG as the input region of the hippocampus plays basic role in processing of information received from sensory organs and other brain's regions. Several studies suggest the role of the DG as a preprocessor of information to send to CA3 [41]. The neural circuit, mechanisms and function of its neurons have been studied intensively in comparison to other sub-regions of the hippocampus. The DG is involved in pattern separation as a basic feature required transforming similar input patterns into subsequently different output patterns [42]. Finally, the DG produces a very sparse coding scheme in which only a small fraction of neurons are active at any time bin of neuronal activity [43] that may be induced by activity of inhibitory neurons in DG [44]. However, the mechanisms of rebalancing of excitatory-inhibitory activity in local circuits in DG as a consequence of different excitatory inputs from EC or network damages to connectivity or impaired inhibitory activities are unknown.

In this work, we aimed to construct a fundamental hippocampus-like artificial memory for neuromorphic technology applications. Such spiking deep neural network uses the structure and functions of hippocampus subregions and their molecular and cellular mechanisms involved in pattern separation in the DG and EC [45]. The action potential of a neuron may create a voltage spike both at the end of the axon (normal propagation) and back through to dendrites, from which much of the original input current originated. This phenomenon is named back-propagation [46, 47] and may trigger retrograde messengers from dendritic site of some neurons. It has been shown that back propagation of action potentials (bAPs) is the principal triggering mechanism of dendritic BDNF secretion occurring during ongoing neuronal activity in neuronal cultures [48]. The simplified model of retrograde messengers plays a critical role in efficiency of the developed system in this study. In the next sections we present the model features and architecture then simulations are explained. The importance of developing such SNN model for biological and artificial applications is presented.

**Method**

The aim of this work is to design a novel spiking neural network that is inspired by rodent's hippocampus equipped with cellular and synaptic mechanisms that can perform pattern separation as the basic information processing in hippocampus using known structure and function of Entorhinal Cortex (EC) and Dentate Gyrus (DG) (**Fig. 1A**).

The artificial system is constructed as a three layered feed-forward neural network (**Fig. 1B**). The sensory layer (**S**) has 200 neurons, and third output layer (**O**) has 1200 neurons. The ratio of neuronal population in these layers has followed the estimated number of neurons in rodents hippocampus in EC and DG regions [41]. The hidden layer has two sets of neurons as excitatory and inhibitory neurons ensembles each 100 neurons. Inhibitory neurons with a constant firing rate equal to 0.4 are activated and send their inputs into **O** layer where pattern separation takes place. The inhibition synaptic plasticity is determined dynamically by retrograde signaling from **O** layer into inhibitory neurons (**I**). The activated neurons in **S** layer are fully connected to excitatory neurons (**E**) in hidden layer and transferred information into **O** layer according to connectivity rate ($Ce$). The excitatory neurons in hidden layer are used to measure Excitation/Inhibition ratio (**E/I**) in the simulations. Connectivity is modeled in the network as the probability of connection of each neuron in a layer to neurons in other layer. There is no connection between neurons in excitation and inhibition neurons in the hidden layer. Neuronal activity of neurons in the first layer is modeled as probabilistic neuron. The neural activity of the second and third layer (output layer) is modeled as Izhikevich model [50] in order to observe sparse activity by changing model's parameters.

$$\frac{dv}{dt} = 0.04\ v^2 + 5v + 140 - u + I(t) \tag{1}$$

$$\frac{du}{dt} = a(bv(t) - u) \tag{2}$$

Where,

$v$ is the membrane potential

$u$ is the parameter that describes the $Na^+$ and $K^+$ ion channel gating

$I$ is the current between synapses

$a$ is the time scale for the parameter $u$ to recover from a spike

$b$ determines the strength of correlation between the membrane potential $v$ and the gating variable $u$

$c$ is the membrane reset potential

$d$ is the post-spike reset for the gating variable u and is equivalent to the slow recovery of the $Na^+$ and $K^+$ concentrations

With the auxiliary after-spike resetting

If $v \geq 30mv$, then $\begin{cases} v \leftarrow c \\ u \leftarrow u + d \end{cases}$

Current into neurons is modeled as (**Equation 3**)

$$\dot{I} = -\frac{I(t)}{\tau_I} + \sum_{t_{pre}} \delta(t - t_{pre}); \quad \tau_I = 20ms \tag{3}$$

The production of retrograde messenger as a consequence of current into postsynaptic neuron is model by **Equation 4.**

$$RM(t) = \frac{t}{\tau_r} exp(-\frac{t}{\tau_r}) \sum_{t_{post}} \delta(t - t_{post}) \quad ; \tau_r = 50ms \tag{4}$$

Where $\sum_{t_{post}} \delta(t - t_{post})$ is the back-propagation signal into dendritic site.

$$\dot{\omega} = -\frac{\omega(t)}{\tau_\omega} + \sum_{t'} RM(t); \quad \tau_\omega = 20ms \tag{5}$$

**Equation 5** shows the dynamic of inhibitory synaptic weight regulated by diffused retrograde messenger induced by back-propagation.

The **E/I** ratio in the simulations is measured as the ratio of average firing rate of excitatory neurons in **S** layer and inhibitory neurons in simulation time. The developed artificial neural system is used in different simulation strategies to adjust the model parameters values to make the system robust to imbalanced Excitation/Inhibition ratio (**E/I** ratio). Such imbalanced **E/I** ratio as a consequence of network damage is very common in biological neural systems [51, 52]. For example, in some cognitive disorders, the imbalance between excitation and inhibition in local circuits involved in sensory and emotional processes has been observed [53-55].

In all simulations, a pair of stimuli trigger layer O as activation of sets of randomly selected neurons (average 40 neurons) with overlap in their pattern between 40 to 70 percent. The objective function is to maximize separation efficiency of the system [44] in response to different E/I ratios. For this purpose, the system parameters is set such that in average firing rate of neurons in layer O equal to 0.4 the separation efficiency is maximized that needs high levels of inhibition intensity in hidden layer. Then the system is checked in other **E/I** ratios. In each experiment, simulation is done 100 times to get average value of separation efficiency.

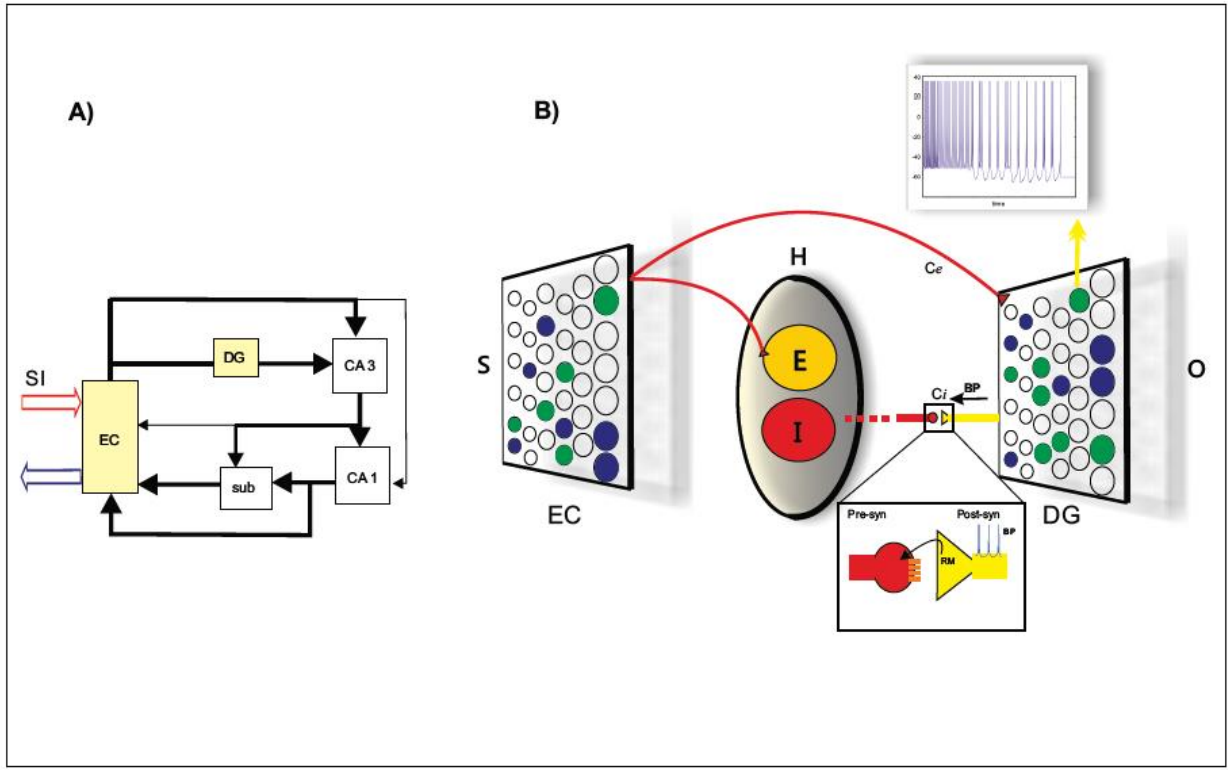

**Figure 1. A. Simplified schematic depiction of hippocampal circuitry.** Sensory inputs (SI) are transferred to EC where the information is transferred into other subregions like DG and CA3. Separation of input patterns is done in EC-DG circuitry. **B. Network architecture.** The spiking neural networked developed as a three layer feed-forward neural network inspired by neural architecture of rodent hippocampus. Sensory layer (**S**) is stimulated as activation of two different sets of neurons (here green and blue neurons) and information is transferred into Hidden layer (**H**) where Excitatory neurons (**E**) receive stimulation. Consequently, inhibitory neurons in **H** send spikes into Output layer (**O**) according to connectivity rates $C_i$. The overlap between activated neurons by two stimuli (green and blue neurons in **O**) determines pattern separation efficiency of the neural system. The neural network is equipped with retrograde signaling from activated neurons in **O** that controls synaptic inhibitory plasticity. *Izhikevich–neuron model* is used in spiking neurons of the neural system. In the model no connections are considered between **E** and **I** neurons.

## Results

The hypothetical mechanism of rebalancing E/I ratio in different conditions is based on diffusion of retrograde messengers from DG neurons into local inhibitory neurons. Fig. 2 shows the dependency of released retrograde messenger induced by backpropagation for different model parameters.

Model's parameters are set to obtain maximum separation efficiency at E/I ratio equal to 0.1 in the absence of backpropagation and retrograde signaling mechanisms in the spiking neural network. To study the role of implemented mechanisms in controlling sparse spiking pattern of the neurons in third layer and their separation efficiency, three levels of E/I ratios are tested to present to the system to measure changes in firing rate of neurons in third layer and synaptic weight of inhibitory neurons (**Fig. 3**). **Fig.3A, B** show the changes in average synaptic weight of inhibitory neurons from initial value equal to 20 for E/I ratio set to 1.2. The system gradually increases the inhibitory weight that induces inhibition into third layer that consequently reduce firing rate of neurons after 500ms. Reducing E/I ratio to 0.8 cause the same effect but with lower intensity (**Fig.3C, D**). The sparse spiking in third layer is observed after 400ms while increase in synaptic weight of inhibitory neurons is raised to about 200. For imbalanced conditions to be less than basic value (=0.5) E/I ratio set to 0.05. Initial inhibitory synaptic weight set to 200. Regarding to low excitatory input (shown by E/I ratio) this value of inhibitory weight prevents spiking of neurons in third layer for a short time while inhibitory weight is reduced to lower values. Then sparse spiking of neurons in third layer is observed (**Fig.3E, F**).

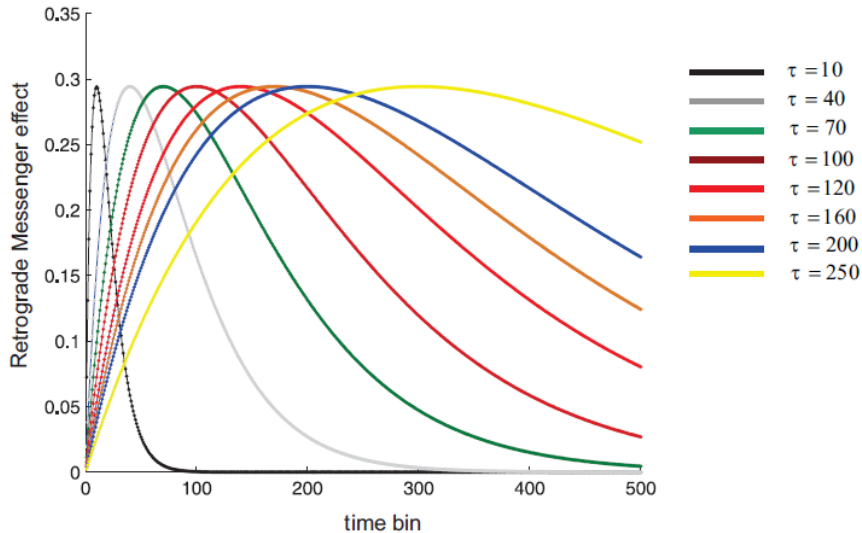

**Figure 2.** **Retrograde messenger level induced by back-propagation signal.** Spiking activity in output neurons leads to back-propagation toward dendrites. Induced retrograde messenger from output neurons into hidden layer induces changes in synaptic plasticity between two layers. A change in messenger level for different model parameter has been shown.

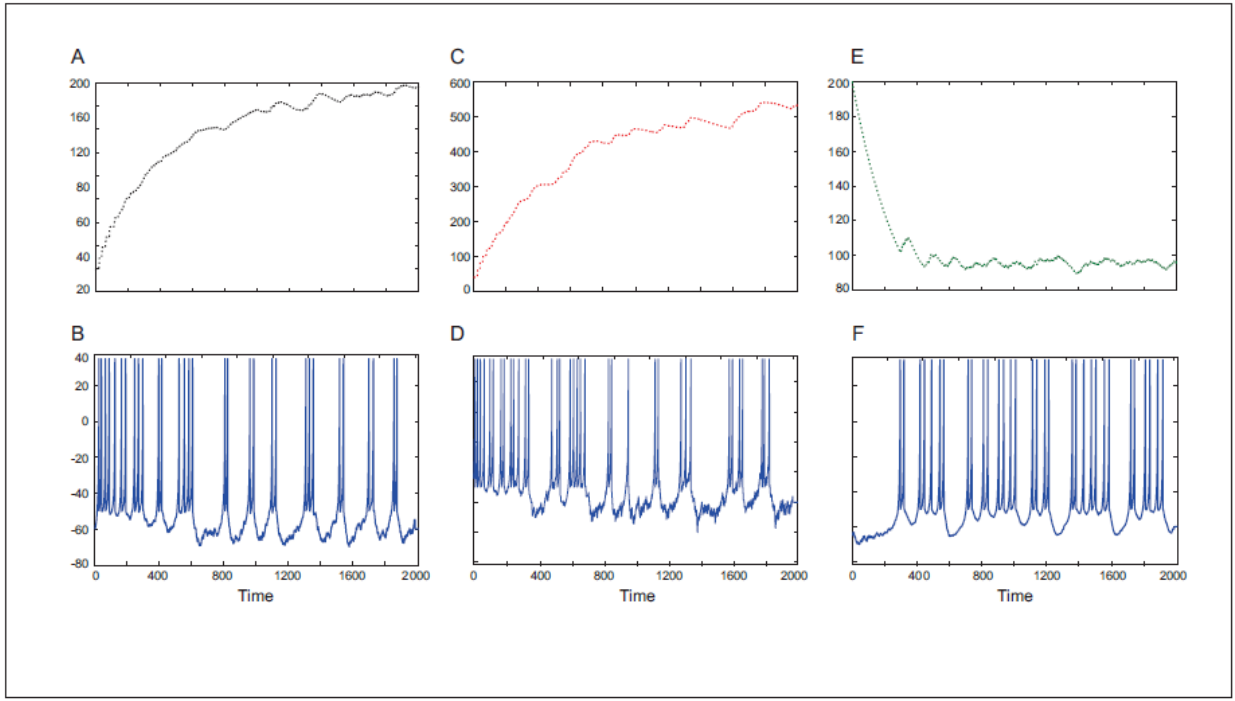

**Figure 3.** Change in average synaptic weight between activated neurons in inhibitory neurons in the hidden layer and the output layer in different E/I ratio equal to 1.2, 0.8, 0.05, respectively (**A, C, E**). Corresponding firing rate of a given neuron in output layer for each E/I ration has shown in **B, D, F**, respectively.

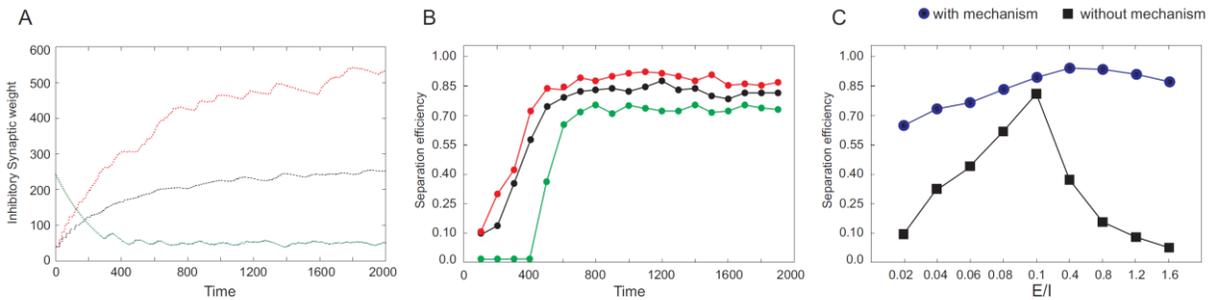

**Figure 4.** **A.** A comparison of changes in synaptic weight of inhibitory neurons for different E/I ratio used in the simulation. In all E/I ratio used in the simulations after a short time synaptic weight has been stable. **B**. The separation efficiency of the systems for different $E/I$ ratio over time. The plot shows increase in efficiency over time when the system has changed synaptic weight of inhibitory neurons. **C.** Separation efficiency of the system for different levels of E/I. The results show the robustness of the system to a change in E/I level induced by different kind of imbalanced excitatory and inhibitory activity of neurons. In the absence of retrograde signaling mechanism, an optimal value for separation efficiency was found and any change to the optimal ratio lead to decrease in efficiency of the system.

**Fig. 4A** shows a comparison between changes in inhibitory synaptic weight for three levels of E/I ratio presented to the system. The simulations demonstrate stability of average inhibitory synaptic weight after 1800ms. Regarding the relation between inhibitory synaptic weight and spiking pattern of neurons in third layer over time, we measure separation efficiency of the system for some synaptic weights (**Fig. 4B**). The results show an increase in separation efficiency over time while inhibitory synaptic weight is stabilized. These simulations motivate to test separation efficiency for larger range of E/I ratios and comparing it to the system without backpropagation and retrograde signaling mechanisms (**Fig. 4C**). The simulations show the robustness of the system equipped with backpropagation and retrograde signaling mechanisms in comparison to the efficiency of the system without this mechanism.

## Discussion

Spiking neural networks mimics the way neurons are connected and communicate in the human brain and have been used in many industrial applications including autonomous robots. However, our limited knowledge on neural circuits underlying cognitive functions is a challenge for developing brain inspired artificial architectures.

The animal brain demonstrate different kind of memories that are dependent on hippocampal and neocortical circuits underlying information processing. Although there is a long way to understand neural mechanisms of cognitive functions, using known cellular and network mechanisms of information flow in hippocampus and neocortex may develop novel artificial systems with capabilities similar to biological memory. For this purpose, novel spiking neural networks inspired by biological information may play critical role in developing artificial architectures [45]. In this work neural architecture of Entorhinal cortex and Dentate Gyrus were used to develop a spiking neural network equipped with a novel hypothetical mechanism to store information efficiently. Although this developed system demonstrates high efficiency in storing information, it can be developed to implement memory retrieval adding CA3 like architecture. In this work, inhibitory layer play critical role in information processing to store input information. The role of inhibitory neurons in hippocampus is not known experimentally but theoretical studies have shown such critical role in pattern separation in DG. In this work, its synaptic weight is dynamically modulated by a hypothetical backpropagation signal from activated neurons in DG that induces diffusion of retrograde messenger from DG neurons into inhibitory neurons that increase the synaptic weight of inhibitory neurons. Such dynamical mechanisms directly control the sparse spiking of DG neurons and are shown in the simulations and so indirectly affect separation efficiency that were measured in the simulations. In the model, a hidden layer was developed to implement different E/I ratios. Imbalanced E/I ratios have been observed in cognitive disorders. In this theoretical study imbalanced E/I ratio is associated with impaired separation efficiency as preliminary stage of episodic memory in absence of hypothetical retrograde signalling induced by theoretical backpropagation in DG neurons. The spiking neural network developed in this study demonstrates high pattern separation efficiency in response to different values of E/I ratios as a consequence of fluctuation in inputs intensity or damages of the number of EC neurons or inhibitory neurons or damages of synapses as information processing units. Therefore, this work presents novel hypothesis on mechanism of robustness to network damages in biological neural networks induced by changes in number of inhibitory neurons or impaired synaptic functions. On the other hand, it can be used in autonomous robots by neuromorphic technology navigating environment to detect stimuli (e.g. visual stimuli) with different levels of stimulation intensity.

**References**

1. GoertzelBen, LianRuiting,Arel Itamar,deGaris Hugo, ChenShuo: A world survey of artificial brain projects, Part ii. Biologically inspired cognitive architectures. Neurocomputing 2010,74:30-49.

2. Hassabis, Demis, Dharshan Kumaran, Christopher Summerfield, and Matthew Botvinick. "Neuroscience-inspired artificial intelligence." Neuron 95, no. 2 (2017): 245-258..

3. Blundell, C., Uria, B., Pritzel, A., Yazhe, L., Ruderman, A., Leibo, J.Z., Rae, J., Wierstra, D., and Hassabis, D. (2016). Model-free episodic control. arXiv, arXiv:160604460.

4. Botvinick, M.M., and Plaut, D.C. (2006). Short-term memory for serial order: a recurrent neural network model. Psychol. Rev. 113, 201–233.

5. Ferbinteanu, Janina, Pamela J. Kennedy, and Matthew L. Shapiro. "Episodic memory—from brain to mind." Hippocampus 16, no. 9 (2006): 691-703.

6. Tulving, E. (2002). Episodic memory: from mind to brain. Annu. Rev. Psychol. 53, 1–25.

7. Squire, L.R., Stark, C.E., and Clark, R.E. (2004). The medial temporal lobe. Annu. Rev. Neurosci. 27, 279–306.

8. Lavenex, Pierre, and David G. Amaral. "Hippocampal-neocortical interaction: A hierarchy of associativity." Hippocampus 10, no. 4 (2000): 420-430.

9. Moser, E. I. & Moser, M. B. One-shot memory in hippocampal CA3 networks. Neuron 38, 147–148 (2003).

10. Weaver, Janelle. "How one-shot learning unfolds in the brain." PLoS biology 13, no. 4 (2015): e1002138.

11. Bannerman, David M., Rolf Sprengel, David J. Sanderson, Stephen B. McHugh, J. Nicholas P. Rawlins, Hannah Monyer, and Peter H. Seeburg. "Hippocampal synaptic plasticity, spatial memory and anxiety." Nature Reviews Neuroscience15, no. 3 (2014): 181.

12. Guderian, Sebastian, Anna M. Dzieciol, David G. Gadian, Sebastian Jentschke, Christian F. Doeller, Neil Burgess, Mortimer Mishkin, and Faraneh Vargha-Khadem. "Hippocampal volume reduction in humans predicts impaired allocentric spatial memory in virtual-reality navigation." Journal of Neuroscience 35, no. 42 (2015): 14123-14131.

13. Spellman, Timothy, Mattia Rigotti, Susanne E. Ahmari, Stefano Fusi, Joseph A. Gogos, and Joshua A. Gordon. "Hippocampal–prefrontal input supports spatial encoding in working memory." Nature 522, no. 7556 (2015): 309.

14. Rothschild, Gideon, Elad Eban, and Loren M. Frank. "A cortical–hippocampal–cortical loop of information processing during memory consolidation." Nature neuroscience 20, no. 2 (2017): 251.

15. Schliebs, Stefan, Haza Nuzly Abdull Hamed, and Nikola Kasabov. "Reservoir-based evolving spiking neural network for spatio-temporal pattern recognition." In International Conference on Neural Information Processing, pp. 160-168. Springer, Berlin, Heidelberg, 2011.

16. Iakymchuk, Taras, Alfredo Rosado-Muñoz, Juan F. Guerrero-Martínez, Manuel Bataller-Mompeán, and Jose V. Francés-Víllora. "Simplified spiking neural network architecture and STDP learning algorithm applied to image classification." EURASIP Journal on Image and Video Processing 2015, no. 1 (2015): 4.

17. Kasabov, Nikola, and Elisa Capecci. "Spiking neural network methodology for modelling, classification and understanding of EEG spatio-temporal data measuring cognitive processes." Information Sciences 294 (2015): 565-575.

18. Tandon, Pulkit, Yash H. Malviya, and Bipin Rajendran. "Efficient and robust spiking neural circuit for navigation inspired by echolocating bats." In Advances in Neural Information Processing Systems, pp. 938-946. 2016.

19. Stromatias, Evangelos, Miguel Soto, Teresa Serrano-Gotarredona, and Bernabé Linares-Barranco. "An event-driven classifier for spiking neural networks fed with synthetic or dynamic vision sensor data." Frontiers in neuroscience 11 (2017): 350.

19. Faghihi, Faramarz, Ahmed A. Moustafa, Ralf Heinrich, and Florentin Wörgötter. "A computational model of conditioning inspired by Drosophila olfactory system." Neural Networks 87 (2017): 96-108.

20. Ge, Chenjie, Nikola Kasabov, Zhi Liu, and Jie Yang. "A spiking neural network model for obstacle avoidance in simulated prosthetic vision." Information Sciences 399 (2017): 30-42.

21. Igarashi, Jun, Osamu Shouno, Tomoki Fukai, and Hiroshi Tsujino. "Real-time simulation of a spiking neural network model of the basal ganglia circuitry using general purpose computing on graphics processing units." Neural Networks 24, no. 9 (2011): 950-960.

22. Cutsuridis, Vassilis, and Thomas Wennekers. "Hippocampus, microcircuits and associative memory." Neural Networks 22, no. 8 (2009): 1120-1128.

23. Kaplan, Bernhard A., and Anders Lansner. "A spiking neural network model of self-organized pattern recognition in the early mammalian olfactory system." Frontiers in neural circuits8 (2014): 5.

24. Chorley, Paul, and Anil K. Seth. "Dopamine-signaled reward predictions generated by competitive excitation and inhibition in a spiking neural network model." Frontiers in computational neuroscience 5 (2011): 21.

25. Tan, Chin Hiong, Huajin Tang, Kay Chen Tan, and Miaolong Yuan. "A hippocampus CA3 spiking neural network model for storage and retrieval of sequential memory." In Cybernetics and Intelligent Systems (CIS), IEEE Conference on, pp. 134-139. IEEE, 2013.

26. Tan, Chin Hiong, Huajin Tang, Eng Yeow Cheu, and Jun Hu. "A computationally efficient associative memory model of hippocampus CA3 by spiking neurons." In Neural Networks (IJCNN), The 2013 International Joint Conference on, pp. 1-8. IEEE, 2013.

27. Isaacson, Jeffry S., and Massimo Scanziani. "How inhibition shapes cortical activity." Neuron 72, no. 2 (2011): 231-243.

28. Dehghani, Nima, Adrien Peyrache, Bartosz Telenczuk, Michel Le Van Quyen, Eric Halgren, Sydney S. Cash, Nicholas G. Hatsopoulos, and Alain Destexhe. "Dynamic balance of excitation and inhibition in human and monkey neocortex." Scientific reports 6 (2016): 23176.

29. Kirkwood, Alfredo. "Balancing excitation and inhibition." Neuron 86, no. 2 (2015): 348-350..

30. O'Donnell, Cian, J. Tiago Goncalves, Carlos Portera-Cailliau, and Terrence J. Sejnowski. "Beyond excitation/inhibition imbalance in multidimensional models of neural circuit changes in brain disorders." Elife 6 (2017): e26724.

31. Wang, Runchun M., Chetan S. Thakur, and Andre van Schaik. "An FPGA-Based Massively Parallel Neuromorphic Cortex Simulator." Frontiers in neuroscience 12 (2018): 213.

32. Cheung, Kit, Simon R. Schultz, and Wayne Luk. "NeuroFlow: a general purpose spiking neural network simulation platform using customizable processors." Frontiers in neuroscience 9 (2016): 516.

33. Duarte RC, Morrison A (2014) Dynamic stability of sequential stimulus representations in adapting neuronal networks. Frontiers in Computational Neuroscience 8: 124. doi: 10.3389/fncom.2014.00124 PMID: 25374534

34. Srinivasa, Narayan, and Youngkwan Cho. "Unsupervised discrimination of patterns in spiking neural networks with excitatory and inhibitory synaptic plasticity." Frontiers in computational neuroscience 8 (2014): 159.

35. Faghihi, Faramarz, and Ahmed A. Moustafa. "The dependence of neuronal encoding efficiency on Hebbian plasticity and homeostatic regulation of neurotransmitter release." Frontiers in cellular neuroscience 9 (2015): 164.

36. Landau, Itamar D., Robert Egger, Vincent J. Dercksen, Marcel Oberlaender, and Haim Sompolinsky. "The impact of structural heterogeneity on excitation-inhibition balance in cortical networks." Neuron 92, no. 5 (2016): 1106-1121.

37. Sprekeler, Henning. "Functional consequences of inhibitory plasticity: homeostasis, the excitation-inhibition balance and beyond." Current opinion in neurobiology 43 (2017): 198-203.

38. Hennequin, Guillaume, Everton J. Agnes, and Tim P. Vogels. "Inhibitory plasticity: balance, control, and codependence." Annual review of neuroscience 40 (2017): 557-579.

39. Vogels, Tim P., Henning Sprekeler, Friedemann Zenke, Claudia Clopath, and Wulfram Gerstner. "Inhibitory plasticity balances excitation and inhibition in sensory pathways and memory networks." Science 334, no. 6062 (2011): 1569-1573.

40. Cline, Hollis. "Synaptogenesis: a balancing act between excitation and inhibition." Current Biology 15, no. 6 (2005): R203-R205.

41. Jonas, Peter, and John Lisman. "Structure, function, and plasticity of hippocampal dentate gyrus microcircuits." Frontiers in neural circuits 8 (2014): 107.

42. Myers, Catherine E., and Helen E. Scharfman. "Pattern separation in the dentate gyrus: a role for the CA3 backprojection." Hippocampus 21, no. 11 (2011): 1190-1215.

43. Diamantaki, Maria, Markus Frey, Philipp Berens, Patricia Preston-Ferrer, and Andrea Burgalossi. "Sparse activity of identified dentate granule cells during spatial exploration." Elife 5 (2016): e20252.

44. Faghihi, Faramarz, and Ahmed A. Moustafa. "A computational model of pattern separation efficiency in the dentate gyrus with implications in schizophrenia." Frontiers in systems neuroscience 9 (2015): 42.

45. Faghihi, Faramarz, and Ahmed A. Moustafa. "Combined computational systems biology and computational neuroscience approaches help develop of future "cognitive developmental robotics"." Frontiers in neurorobotics 11 (2017): 63.

46. Grewe, Benjamin F., Audrey Bonnan, and Andreas Frick. "Back-propagation of physiological action potential output in dendrites of slender-tufted L5A pyramidal neurons.", Frontiers in cellular neuroscience 4 (2010).

47. Yu, Yuguo, Yousheng Shu, and David A. McCormick. "Cortical action potential backpropagation explains spike threshold variability and rapid-onset kinetics." Journal of Neuroscience 28, no. 29 (2008): 7260-7272.

48. Kuczewski, Nicola, Christophe Porcher, Nadine Ferrand, Hervé Fiorentino, Christophe Pellegrino, Richard Kolarow, Volkmar Lessmann, Igor Medina, and Jean-Luc Gaiarsa. "Backpropagating action potentials trigger dendritic release of BDNF during spontaneous network activity." Journal of Neuroscience 28, no. 27 (2008): 7013-7023.

49. Jinno, Shozo, and Toshio Kosaka. "Stereological estimation of numerical densities of glutamatergic principal neurons in the mouse hippocampus." Hippocampus 20, no. 7 (2010): 829-840.

50. Izhikevich, Eugene M. "Simple model of spiking neurons." IEEE Transactions on neural networks 14, no. 6 (2003): 1569-1572.

51. Lee, Eunee, Jiseok Lee, and Eunjoon Kim. "Excitation/inhibition imbalance in animal models of autism spectrum disorders." Biological psychiatry 81, no. 10 (2017): 838-847.

52. Yizhar, Ofer, Lief E. Fenno, Matthias Prigge, Franziska Schneider, Thomas J. Davidson, Daniel J. O'shea, Vikaas S. Sohal et al. "Neocortical excitation/inhibition balance in information processing and social dysfunction." Nature 477, no. 7363 (2011): 171.

53. Tatti, Roberta, Melissa S. Haley, Olivia K. Swanson, Tenzin Tselha, and Arianna Maffei. "Neurophysiology and regulation of the balance between excitation and inhibition in neocortical circuits." Biological psychiatry 81, no. 10 (2017): 821-831.

54. Marín, Oscar. "Interneuron dysfunction in psychiatric disorders." Nature Reviews Neuroscience 13, no. 2 (2012): 107.